\documentclass[aps,prl,twocolumn,amsmath,amssymb,nofootinbib,showpacs,superscriptaddress]{revtex4-1}
\usepackage[english]{babel}
\usepackage{latexsym}
\usepackage{graphics}
\usepackage{epsfig}
\usepackage{color}
\usepackage{bm}
\usepackage{amsmath}
\usepackage{amssymb}
\usepackage{tikz}  \usetikzlibrary{patterns}

\begin{document}

\title{Finite temperature disordered bosons in two dimensions}

\author{G. Bertoli}
\affiliation{LPTMS, CNRS, Univ. Paris-Sud, Universit\'e Paris-Saclay, Orsay 91405, France}

\author{V.P. Michal}
\affiliation{QuTech and Kavli Institute of Nanoscience, TU Delft, 2600 GA Delft, The Netherlands }

\author{B.L. Altshuler}
\affiliation{Physics Department, Columbia University, 538 West 120th Street, New York, New York 10027, USA}

\author{G.V. Shlyapnikov}
\affiliation{LPTMS, CNRS, Univ. Paris-Sud, Universit\'e Paris-Saclay, Orsay 91405, France}
\affiliation{SPEC, CEA, CNRS, Universit\'e Paris-Saclay, CEA Saclay, Gif sur Yvette 91191, France}
\affiliation{Russian Quantum Center, Skolkovo, Moscow Region 143025, Russia}
\affiliation{\mbox{Van der Waals-Zeeman Institute, University of Amsterdam, Science Park 904, 1098 XH Amsterdam, The Netherlands}}
\affiliation{State Key Laboratory of Magnetic Resonance and Atomic and Molecular Physics,
Wuhan Institute of Physics and Mathematics, Chinese Academy of Sciences, Wuhan 430071, China}

\date{\today}
\begin{abstract}
We study phase transitions in a two dimensional weakly interacting Bose gas in a random potential at finite temperatures. We identify superfluid, normal fluid, and insulator phases and construct the phase diagram. At $T=0$ one has a tricritical point where the three phases coexist. The truncation of the energy distribution at the trap barrier, which is a generic phenomenon in cold atom systems, limits the growth of the localization length and in contrast to the thermodynamic limit the insulator phase is present at any temperature. 
\end{abstract}
\maketitle

After almost 60 years since its discovery \cite{Anderson}, the concept of Anderson localization (AL) of a quantum particle by a quenched disorder remains an extremely active direction of research \cite{Abrahams}. To a large extent, this is due to a subtle problem of the effect of interaction between particles on the localization \cite{Fleishman, Altshuler, Fisher,Basko}. It has been demonstrated that interacting particles can undergo many-body localization-delocalization transition (MBLDT), that is the transition from insulator to fluid state \cite{Basko}. A new wave of interest to this problem was inspired by the observation of AL in dilute quasi-one-dimensional clouds of cold bosonic atoms with a negligible interaction \cite{Billy, Roati}. Presently, the studies of ultracold interacting atoms form a rapidly growing domain in the physics of disordered quantum systems \cite{Sanchez-Palencia}. Although the first observations of MBLDT in these systems have been reported \cite{Tanzi,Schreiber,Choi}, many features of MBLDT remain unexplored, especially in higher than one dimension. In contrast to the one-dimensional (1D) case where at any non-zero temperature, $T>0$, only normal fluid and insulator (glass) phases are possible, in two dimensions (2D) the phase diagram contains one more phase. Two-dimensional bosons undergo the Berezinskii-Kosterlitz-Thouless (BKT) transition \cite{Ber1971,Kosterlitz} and form an algebraic superfluid below a critical temperature $T_{BKT}$. While a number of studies \cite{Falco,Carleo,{Nandkishore}} was devoted to evaluating the critical disorder strength either for the MBLDT at zero temperature or for the BKT transition, the full finite temperature phase diagram of such a system to the best of our knowledge has never been published \cite{note1}.

In this Letter we construct the phase diagram of 2D weakly interacting bosons subject to a static random potential. The diagram is displayed in Fig.1 in terms of $T$ and $\epsilon_\ast$, where the energy scale $\epsilon_\ast$ characterizes the disorder strength. It turns out that there are two temperature dependent critical values of disorder: $\epsilon_\ast^{BKT}(T)$ and $\epsilon_\ast^{MBL}(T)$, i.e. two separatrices in Fig. 1 \cite{note1}. The first one separates the normal fluid from the superfluid phase and it shows the suppression of superfluidity by the disorder. Since superfluidity disappears at $T>T_{BKT}$ even without disorder, we have $\epsilon_\ast^{BKT}(T\geq T_{BKT})=0$. For sufficiently strong disorder, $\epsilon_\ast>\epsilon_\ast^{BKT}(0)$, the superfluid regime is absent even at $T=0$. The second separatrix is the MBLDT curve. The region $\epsilon_\ast>\epsilon_\ast^{MBL}(T)$ corresponds to the insulator (glass) phase, which undergoes a transition to the normal fluid as the disorder is reduced to below $\epsilon_\ast^{MBL}(T)$. 

The important property of 2D weakly interacting disordered bosons is the instability of the normal fluid at $T=0$ with respect to a transition either to the superfluid or to the insulator regime. Accordingly, one has
\begin{equation}     \label{eps0}
\epsilon_\ast^{MBL}(0)=\epsilon_\ast^{BKT}(0).
\end{equation}
This means that the point $T=0, \epsilon_\ast=\epsilon_\ast^{MBL}(0)$ is a tricritical point, where the three phases coexist \cite{note1}.
 
In terms of field operators $\hat\Psi({\bf r})$, the Hamiltonian of 2D interacting disordered bosons reads:
\begin{align}
&\hat H = \int d^2r\left(-\hat\Psi^\dagger(\textbf{r})\frac{\hbar^2}{2m}\nabla^2\hat\Psi(\textbf{r})
+g\hat\Psi^\dagger(\textbf{r})\hat\Psi^\dagger(\textbf{r})\hat\Psi(\textbf{r})\hat\Psi(\textbf{r})\right.\nonumber\\
&\label{eq:hamiltonian}\left.\qquad\qquad\qquad+\hat\Psi^\dagger(\textbf{r})U(\textbf{r})\hat\Psi(\textbf{r})\right).
\end{align}
The first term is the kinetic energy of particles ($m$ is the particle mass), and the second term (denoted below as $H_{int}$) describes a contact interaction between them, characterized by the coupling constant $g>0$. The third term represents the effect of the random potential $U(\bf{r})$. We assume that $U({\bf r})$ is a Gaussian short-range potential with zero mean, correlation length $\sigma$ and amplitude $U_0$ such that $U_0\ll \hbar^2/m\sigma^2$. The only disorder-related length and energy scales are known to be \cite{Lifshitz,Zittartz}
\begin{equation}
\zeta_\ast = \sqrt{\frac{2e^2}{\pi}}\frac{\hbar^2}{mU_0\sigma};\qquad\epsilon_\ast = \frac{m U_0^2\sigma^2}{\pi\hbar^2}.
\end{equation}

In the absence of disorder the density of states (DoS) for 2D bosons in the continuum is energy independent, $\rho_0 = m/2\pi\hbar^2$. The random potential creates negative energy states, which form the so-called Lifshitz tails: the DoS decays exponentially as the absolute value of the energy increases \cite{Lifshitz,Zittartz}. Below we omit these states. For positive energies $\epsilon\gg\epsilon_\ast$ and even for $|\epsilon|\lesssim\epsilon_\ast$ the effect of the disorder is limited and $\rho(\epsilon)\simeq\rho_0$ is a good approximation.  

In two dimensions all single particle states are localized. The localization length $\zeta$ increases exponentially with the particle energy for $\epsilon>\epsilon_\ast$ \cite{Lee}:
\begin{equation}\label{eq:locLengthEn}
\zeta(\epsilon) = \frac{\zeta_\ast}{e}\sqrt{\frac{\epsilon}{\epsilon_\ast}}e^{\epsilon/\epsilon_\ast};\,\,\,\,\,\epsilon\gg\epsilon_\ast,
\end{equation}
which was, in particular, observed in atomic kicked rotor experiments \cite{Manai}. At energies $|\epsilon|\lesssim\epsilon_\ast$ one can neglect the energy dependence of $\zeta$ and approximate the localization length as $\zeta(\epsilon)\approx\zeta_\ast$.

We consider the weakly interacting regime, where the degeneracy temperature $T_d=2\pi\hbar^2n/m$ greatly exceeds the mean interaction energy per particle $ng$, with $n$ being the mean density. Thus, there is a small parameter 
\begin{equation}    \label{sp}
\frac{ng}{T_d}=\frac{mg}{2\pi\hbar^2}\ll 1.
\end{equation}
We also assume that the disorder is weak, so that 
\begin{equation}     \label{wd}
\epsilon_\ast\ll T_d.
\end{equation}

In order to estimate the critical disorder $\epsilon_\ast^{MBL}$ at a given $g$, we employ the method developed in Refs. \cite{Basko,Aleiner}. Namely, we consider a particular one-particle localized state $|\alpha\rangle$ and evaluate the probability $P_{\alpha}$ that there exist three other states $|\beta\rangle, |\alpha'\rangle,|\beta'\rangle$ such that the two-particle states $|\alpha,\beta\rangle$ and $|\alpha',\beta'\rangle$ are at resonance. This means that the matrix element of the interaction $\langle\alpha',\beta'|H_{int}|\alpha,\beta\rangle$ exceeds the energy mismatch $\Delta^{\alpha'\beta'}_{\alpha\beta}=|\epsilon_{\alpha}+\epsilon_{\beta}-\epsilon_{\alpha'}-\epsilon_{\beta'}|$, where $\epsilon_{\alpha},\epsilon_{\beta},\epsilon_{\alpha'},\epsilon_{\beta'}$ are one-particle energies. The MBLDT occurs when $P_{\alpha}$ becomes close to unity.  

The matrix elements of the interaction are small unless the energies $\epsilon_{\alpha},\epsilon_{\beta},\epsilon_{\alpha'},\epsilon_{\beta'}$ are almost equal pairwise, e.g. $\epsilon_{\alpha}\approx\epsilon_{\alpha'}$ and $\epsilon_{\beta}\approx\epsilon_{\beta'}$. Then we have (see \cite{Basko,Aleiner}):
\begin{equation}    \label{me}
\langle\alpha',\beta'|H_{int}|\alpha,\beta\rangle\simeq\frac{gN_{\beta}}{{\rm max}(\zeta_{\alpha}^2,\zeta_{\beta}^2)},
\end{equation}
where $\zeta_{\alpha,\beta}\equiv\zeta(\epsilon_{\alpha,\beta})$, and $N_{\beta}$ is the occupation number for the state $|\beta\rangle$. 

For $|\alpha\rangle$ and $\alpha'\rangle$ being nearest neighbors in energy the energy mismatch is $\Delta^{\alpha'\beta'}_{\alpha\beta}=|\delta_{\alpha}+\delta_{\beta}|$, where $\delta_{\alpha}$ is the level spacing between the states on the length scale close to $\zeta_{\alpha}$. The mismatch can thus be estimated as
\begin{equation}      \label{mis}
\Delta^{\alpha'\beta'}_{\alpha\beta}\simeq {\rm max}(\delta_{\alpha},\delta_{\beta})=\frac{1}{{\rm min}(\rho_{\alpha}\zeta_{\alpha}^2,\rho_{\beta}\zeta_{\beta}^2)},
\end{equation}
and $\langle\alpha',\beta'|H_{int}|\alpha,\beta\rangle$ exceeds $\Delta^{\alpha'\beta'}_{\alpha\beta}$ for given $|\alpha\rangle, |\beta\rangle, |\alpha'\rangle,|\beta'\rangle$ with the probability 
\begin{equation}
P^{\alpha'\beta'}_{\alpha\beta}=\frac{\langle\alpha',\beta'|H_{int}|\alpha,\beta\rangle}{\Delta^{\alpha'\beta'}_{\alpha\beta}}.
\end{equation}
The quantity $P_{\alpha}$ is the sum of  $P^{\alpha'\beta'}_{\alpha\beta}$ over $\beta,\alpha',\beta'$, and the MBLDT criterion takes the form (see \cite{Basko,Aleiner}):
\begin{equation}     \label{MBLDT1}
g_c\sum_{\beta}N_{\beta}\frac{{\rm min}(\rho_{\alpha}\zeta_{\alpha}^2,\rho_{\beta}\zeta_{\beta}^2)}{{\rm max}(\zeta_{\alpha}^2,\zeta_{\beta}^2)}=C,
\end{equation}
where $C$ is a model-dependent coefficient of order unity. However, varying $C$ does not affect the main conclusions of this Letter and below we use $C=1$ (see Supplemental Material).

Omitting Lifshitz tails we replace the summation over $\beta$ in Eq.(\ref{MBLDT1}) by the integration over $\epsilon_{\beta}$ with the lower limit $-\epsilon_\ast$. Taking into account that the DoS is energy independent and equal to $\rho_0$ we transform equation (\ref{MBLDT1}) to
\begin{equation}\label{eq:MBLCriterion}
g(\epsilon_{\alpha})\rho_0^2 \left(\frac{1}{\zeta^2(\epsilon_{\alpha})}\int_{-|\epsilon_\ast|}^{\epsilon_{\alpha}}\!\!\!\!\!\!d\epsilon N_{\epsilon}\zeta^4(\epsilon)+\zeta^2(\epsilon_{\alpha})\int_{\epsilon_\alpha}^\infty\!\!\!\!\!\!d\epsilon N_\epsilon\right)\!=1.
\end{equation}

The coupling strength $g$ as determined by Eq.(\ref{eq:MBLCriterion}) depends on $\epsilon_{\alpha}$. The latter should be chosen such that it minimizes $g(\epsilon_{\alpha})$, and the critical coupling is $g_c={\rm min}\{g(\epsilon_{\alpha})\}$. 
The occupation numbers $N_{\epsilon}$ depend on the chemical potential $\mu$. Hence, Eq. (\ref{eq:MBLCriterion}) should be complemented with the number equation, which relates $\mu$ and the density $n$:
\begin{equation}     \label{number}
\int_{-|\epsilon_{\ast}|}^{\infty}\rho_0 N_{\epsilon}d\epsilon=n.
\end{equation} 
On the insulator side we have:
\begin{equation}     \label{Negen}
N_{\epsilon}=\left[\exp\left(\frac{\epsilon-\mu+N_{\epsilon}g/\zeta^2(\epsilon)}{T}\right)-1\right]^{-1}.
\end{equation}
For $N_{\epsilon}\gg 1$, i.e. for $T\gg(\epsilon-\mu)$ at $\epsilon>\mu$, we expand the exponent in Eq.(\ref{Negen}) and obtain (see \cite{Michal}):
\begin {equation}     \label{Nelarge}
N_{\epsilon}=\frac{\zeta^2(\epsilon)}{2g}\left(\mu-\epsilon +\sqrt{(\mu-\epsilon)^2+\frac{4Tg}{\zeta^2(\epsilon)}}\right).
\end{equation}

In what follows, we refer the reader to the Supplemental Material for the calculation details, and show only the main results.

At zero temperature Eq.(\ref{Nelarge}) gives 
\begin{equation}    \label{N0}
N_{\epsilon}=\frac{\zeta^2(\epsilon)(\mu-\epsilon)}{g}\theta(\mu-\epsilon),
\end{equation}
where $\theta(\mu-\epsilon)$ is the theta-function. Combining equations (\ref{N0}), (\ref{number}), and (\ref{eq:MBLCriterion}) we find that $g_c$ is minimized at $\epsilon_{\alpha}=1.93\epsilon_\ast$. The resulting critical disorder as a function of $g$ is
\begin{equation}   \label{epsilon0MBL}
\epsilon_\ast^{MBL}(0)=0.54 ng,
\end{equation}
with the corresponding chemical potential $\mu=1.21ng$. 
The result of Eq. \eqref{epsilon0MBL} is consistent with those obtained from the analysis of tunneling between bosonic lakes \cite{Falco}. 

Corrections to the zero temperature result \eqref{epsilon0MBL} are small as long as $T\ll\epsilon_\ast$. For calculating these corrections one integrates over $\epsilon$ in Eqs. (\ref{eq:MBLCriterion}) and (\ref{number}). This gives the following critical disorder:
\begin{equation}    \label{epsilonTMBL}
\epsilon_\ast^{MBL}(T)=\epsilon_\ast^{MBL}(0)\left[1+0.66\frac{T}{T_d}\ln\left(0.09\frac{T_d}{\epsilon_\ast^{MBL}(0)}\right)\right].
\end{equation}

Exponential increase of the localization length with the particle energy supports delocalization. In the thermodynamic limit, as discussed in Ref. \cite{Nandkishore}, this leads to the disappearance of the insulating phase at temperatures $T>\epsilon_*/2$. However, for realistic systems of cold bosonic atoms the energy distribution is truncated at sufficiently large energy. Indeed, in the process of evaporative cooling, atoms with energies above the trap barrier immediately leave the trap, and the distribution function $N_\epsilon$ is effectively truncated at a finite energy barrier $\epsilon_b$. Typical values of this energy for evaporative cooling to temperatures $T\gtrsim ng$ are equal to $\eta T$, where $\eta$ ranges from 5 to 8 (see, e.g. \cite{walraven,ketterle}). For cooling to temperatures $T\lesssim ng$ the value of the energy barrier can be written as $\epsilon_b=ng+\eta T$ \cite{chang2016}. Below we use $\eta=5$ and, in order to match the zero temperature result, we truncate $N_{\epsilon}$ at $\epsilon_b=1.21ng+\eta T$. Increasing $\eta$ up to 8 has little effect on the MBLDT transition line $\epsilon_\ast^{MBL}(T)$.

The truncation of the energy distribution practically does not influence the results at $T\ll \epsilon_\ast$ and thus equation \eqref{epsilonTMBL} remains valid. However, at higher temperatures the truncation strongly limits the growth of the localization length, and the critical coupling $g_c$ remains finite even for $T>\epsilon_*/2$, i.e. the insulator phase survives. In this case the expression for the critical disorder, valid for $T\ll\epsilon_b$, is: 
\begin{equation}
\epsilon_*^{MBL}(T)=\frac{2\epsilon_b}{\ln \left(4 \pi^3 T_d e^{\epsilon_b/T}/ng \right)-\ln\ln\left(4 \pi^3 T_d e^{\epsilon_b/T}/ng \right)}. \label{epsilonbarrier}
\end{equation}
Equations \eqref{epsilonTMBL}-\eqref{epsilonbarrier} are in good agreement with the numerical solution of Eqs. \eqref{eq:MBLCriterion}-\eqref{Negen}.

Actually, the distribution function $N_{\epsilon}$ does not abruptly go to zero at $\epsilon=\epsilon_b$. It undergoes a smooth, although quite sharp, decrease to zero near $\epsilon_b$ \cite{walraven,ketterle}. The disorder potential introduces an additional smoothness of $N_{\epsilon}$. However, for a weak disorder, the disorder-induced increase of the energy interval near $\epsilon_b$, in which the distribution function goes to zero, is significantly smaller than $U_0^2/\epsilon_b$, and is only a fraction of $\epsilon_*$ for realistic parameters of the system. Our calculations show that this does not change the result of equations \eqref{epsilonTMBL}-\eqref{epsilonbarrier} by more than a few percent.

In the recent paper \cite{Gornyi} it was claimed that many-body localization is prevented in continuum systems. The conclusion was based on the exchange of energy between highly energetic particles and states with typical energies. Without entering the discussion of collisional integrals, we simply note that the truncation of the distribution function (which should clearly emerge after several collision times \cite{walraven}) means that such high-energy particles are not there to induce delocalization. 

It is worth noting that MBLDT can be measured for typical values of disorder, temperature, and density of 2D trapped bosonic atoms. The most promising is the situation where all single-particle states are localized. For example, at densities $n\simeq 10^7$ cm$^{-2}$ of $^{7}$Li atoms  the degeneracy temperature is $T_d\simeq 50$ nK. For the amplitude of the disorder potential,  $U_0 = 35 \text{ nK}$, and correlation length $\sigma\simeq 1.4 \mu$m, we have $\zeta_*\approx 3\mu$m and $\epsilon_*\approx 11.5$ nK. Considering temperatures $T\sim 10$ nK, for barrier energies $\epsilon_b\approx 44 $ nK, the localization length at maximum particle energies can be estimated as $\sim 100\mu$m. The size of the system can be significantly larger, so that all single-particle states are really localized. The MBLDT can be identified by opening the trap. If most of the sample is in the insulator phase, then only a small fraction of particles will escape and the size of the remaining cloud will increase by an amount of the order of the localization length. On the contrary, if most of the sample is in the fluid phase, switching off the trap will lead to the expansion of the major part of the cloud. The MBLDT can be also identified {\it in situ} by measuring the dynamical structure factor with the use of the Bragg spectroscopy, the method employed to distinguish between the superfluid and Mott insulator phases of lattice atomic systems  (see, e.g. \cite{clement,ernst}).

We now start our discussion of the BKT transition between the normal fluid and superfluid phases with the high temperature regime, $T\gg ng$. In the superfluid phase we assume that density fluctuations are small and the Bogoliubov approach remains valid in the presence of disorder. Following Refs. \cite{Huang, Meng} we consider a weak disorder, $\epsilon_{\ast}\ll ng$, and rely on the Hamiltonian ${\cal H}=H_0+\int U({\bf r})\delta n({\bf r})d^2r$, where $H_0$ is the standard Bogoliubov Hamiltonian in the density-phase representation, while the second term describes the interaction of the density fluctuations $\delta n({\bf r})$ with disorder. Diagonalizing $H_0$ and using the known relation for the density fluctuations we have:
\begin{equation}    \label{calH}
{\cal H}=\sum_{\bf k}\hbar\omega_k b^{\dagger}_{\bf k}b_{\bf k}+\sum_{\bf k}nU_{\bf k}(b_{\bf k}+b^{\dagger}_{-\bf k})\sqrt{E_k/\hbar\omega_k}.
\end{equation}
Here $n$ is the mean density, $b_{\bf k}$ and $\hbar\omega_k=\sqrt{E_k^2+2ngE_k}$ are the operators and energies of Bogoliubov excitations with momentum ${\bf k}$, $E_k=\hbar^2k^2/2m$ is the free particle kinetic energy, and $U_{\bf k}$ is the Fourier transform of the disorder potential $U({\bf r})$. For the normal density we then have \cite{Meng}:
\begin{equation}    \label{nf}
n_f=\frac{1}{2}n\!\!\int\!\!\frac{\langle U^*_kU_k\rangle}{(ng+E_k/2)^2}\frac{d^2k}{(2\pi)^2}-\!\!\int\!\! E_k\frac{\partial N_k}{\partial\hbar\omega_k}\frac{d^2k}{(2\pi)^2},
\end{equation}
where we put the normalization volume equal to unity. The result of the integration in the first term of Eq.(\ref{nf}) depends on the correlation function of the disorder. For $\langle U({\bf r})U({\bf r}')\rangle=U_0\delta[({\bf r}-{\bf r}')/\sigma]$ we have $\langle U^*_kU_k\rangle=U_0^2\sigma^2$ and at temperatures $T\gg ng$ equation (\ref{nf}) yields:
\begin{equation}    \label{nfTg}
n_f=\frac{\epsilon_{\ast}}{2g}+\frac{mT}{2\pi\hbar^2}\ln{\frac{T}{ng}};\,\,\,\,\,\,\,\,\,T\gg ng.
\end{equation}

The Bogoliubov approach works well in the superfluid phase, but it does not allow one to determine the exact value of the BKT transition temperature $T_{BKT}$. At this temperature the superfluid density $n_s$ undergoes a jump, and just below $T_{BKT}$ the superfluid density satisfies the Nelson-Kosterlitz relation \cite{Nelson}:
\begin{equation}     \label{nsTKT}
n_s(T_{BKT})=\frac{2m}{\pi\hbar^2}T_{BKT}.
\end{equation}
For $\epsilon_*\ll ng$, the superfluid density $n_s$ next to the BKT transition point is sufficiently large. Hence, it is possible to complement the Nelson-Kosterlitz relation with the expression for $n_s$ from Bogoliubov theory. From equations (\ref{nfTg}) and (\ref{nsTKT}) we obtain a relation for the critical disorder of the BKT transition:
\begin{equation}     \label{gKT}
\epsilon_{\ast}^{BKT}(T)=2ng\left[1-\frac{T}{T_d}\ln\left(e^4\frac{T}{ng}\right)\right].
\end{equation}

In the absence of disorder, the most precise value of $T_{BKT}$ was obtained in Ref. \cite{Prokof'ev} by Monte Carlo simulations: $T_{BKT} = T_d/\ln(\xi T_d/ng)$ with $\xi\simeq 380/2\pi\simeq 60$. In the limit $\epsilon_\ast \to 0$, Eq. (\ref{gKT}) gives $T_{BKT}\simeq T_d /(\ln(e^4T_d/ng)+O(\ln\ln T_d/ng))$. Therefore, $T_{BKT}$ with $n_s$ following from the Bogoliubov approach is close to the exact value of Ref. \cite{Prokof'ev}. This justifies the validity of our method.
For the Gaussian disorder correlation function, Eqs. (\ref{nf}) and (\ref{nsTKT}) lead to critical values of the disorder versus $(T_{BKT}-T)$, which for low disorder agree within 20\% with Monte Carlo calculations \cite{Carleo}.

The employed Bogoliubov approach has to be corrected when $ng$ is approaching $\epsilon_\ast$. In this case the first term of Eqs. \eqref{nf} and \eqref{nfTg} should be complemented by the contribution of higher order diagrams. This can be done by keeping nonlinear (in $b_{\bf k}$) interactions between atoms and random fields in the Hamiltonian \eqref{calH}, as it was done in the three-dimensional case in Ref. \cite{Yukalov}. Instead of equation (\ref{gKT}) we then have:
\begin{equation}     \label{epsilonBKT}
\frac{\epsilon_{\ast}^{BKT}}{2ng}=\left[1-\frac{T}{T_d}\ln\left(e^4\frac{T}{ng}\right)\right]f\left(\frac{\epsilon_\ast^{BKT}}{2ng}\right),
\end{equation} 
where the function $f(x)$ is of order unity.

The BKT transition has been measured in ultracold atomic gases for clean harmonically trapped systems \cite{bktMeasurements}. In the presence of disorder, coherence properties near the BKT superfluid transition \cite{disorderMeasurements} and the resistance for a strongly interacting gas \cite{Krinner} have been studied experimentally. We thus believe that an experimental validation of our results is possible in both harmonically trapped and uniform (box) confining potentials. The 2D Bose gas in a box potential has been created in a number of experiments \cite{boxMeasurements}, in particular with a tunable interaction strength \cite{Ville}, and realistic proposals of how to identify the BKT transition in this system have been made \cite{Mathey}.

Returning to the phase diagram we should admit that close to the tricritical point equations (\ref{epsilon0MBL}) and (\ref{epsilonBKT}) can give only estimates rather than exact values of the critical disorder strengths $\epsilon_\ast^{MBL}$ and $\epsilon_\ast^{BKT}$ (because of not exactly known values of the constant $C$ and function $f$). In particular, in Fig.1 we took $C=1$ and put $f=0.27$ for $\epsilon_\ast=0.54ng$. However, we argue that the identity (\ref{eps0}) holds irrespective of the precision of our approximations and now we present the proof of this identity \cite{note1}.

First of all, $\epsilon_\ast^{BKT}(0)$ cannot exceed $\epsilon_\ast^{MBL}(0)$. As it is explained in detail in the Supplemental Material, such a situation is not possible because the critical line for MBLDT is monotonically increasing, whereas the critical line for the BKT transition is monotonically decreasing. Whereas elementary excitations are extended in the superfluid, in the insulator they are localized by definition. Thus the localization length diverges when $\epsilon_\ast$ approaches $\epsilon_\ast^{MBL}(0)+0$. However, at any fixed disorder $\epsilon_*>\epsilon_*^{MBL}(0)$, the elementary excitations undergo many-body delocalization with increasing temperature. The critical temperature tends to zero as the localization length diverges, i.e. at arbitrary low finite temperatures there will be a range of disorder strengths corresponding to a normal fluid. 

On the other hand, $\epsilon_\ast^{MBL}(0)$ can not exceed $\epsilon_\ast^{BKT}(0)$ either. Indeed, this would mean that the normal fluid is realized at $T=0$ in a certain range of $\epsilon_\ast$, i.e. elementary excitations are extended. However, as follows from the theory of weak localization (see, e.g. \cite{Lee}) in 2D this is impossible for a non-superfluid state. At $T=0$ the normal fluid is unstable with respect to the transition either to an insulator or to a superfluid, depending on the disorder. 

\begin{figure}[h!]
    \centering
        \includegraphics[width=0.48\textwidth]{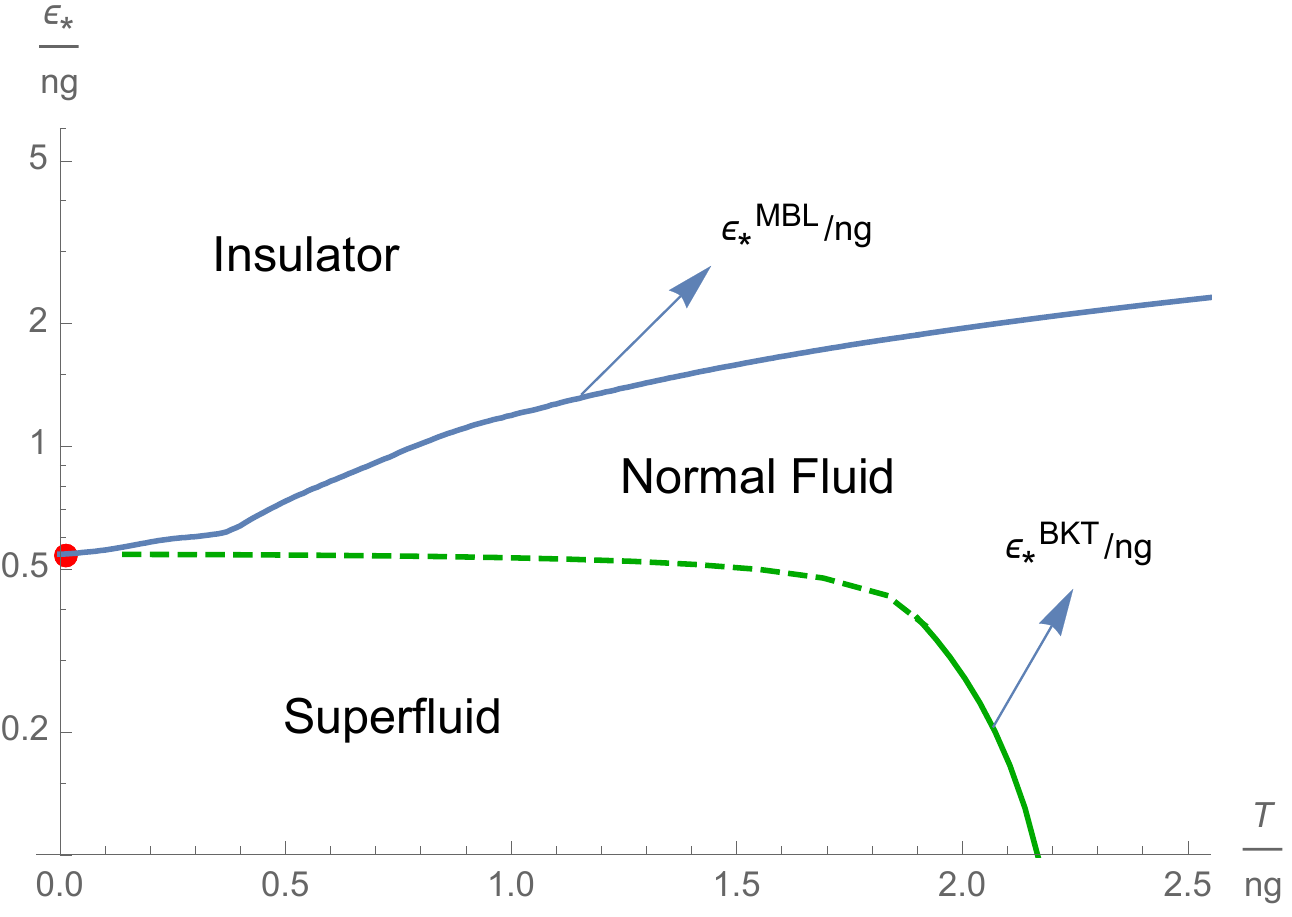}
\caption
{Phase diagram for 2D weakly interacting disordered bosons in terms of the dimensionless disorder strength $\epsilon_{\ast}/ng$ and temperature $T/ng$ for $T_d/ng=11$, with $C=1$ and $f(0.54)=0.27$. The MBLDT border between the insulator and normal fluid follows almost a horizontal line $\epsilon_{\ast}/ng \simeq 0.54$ until the disorder approaches $ \epsilon_{\ast}/ng=2 T / ng$. The line of the MBLDT is obtained with the distribution function truncated at $\epsilon_b=1.21ng+5T$. The solid part of the normal fluid-superfluid line is the result of equation \eqref{gKT}, and the dashed part is our expectation of how it continues at $T\lesssim ng$ until it reaches the tricritical point at $T=0$ (red point).
}
\label{fig:phaseD}
\end{figure}

We thus arrive at the phase diagram of Fig.1 with $\epsilon_\ast^{MBL}(0)=\epsilon_\ast^{BKT}(0)$, which should be valid as long as there exist only three phases: insulator, normal fluid, and superfluid. At low temperatures all phase transitions occur at the coupling strength $ng\sim\epsilon_{\ast}$. In this respect it is worth noting that in the recent experiment on disordered 2D lattice bosons \cite{Choi} it was observed that MBLDT happens when the interaction energy and the characteristic disorder are of the same order of magnitude. 

One may think of a possible alternative to the phase diagram of Fig.1. A phase with non-ergodic but extended eigenstates (non-ergodic phase; see \cite{BK2014} for discussion of such states) can take place in the vicinity of the tricritical point. Detailed discussion of such a possibility goes beyond the scope of the present paper.

We acknowledge discussions with Laurent Sanchez-Palencia, Markus Holzmann and Giuseppe Carleo, and we are especially grateful to Igor Aleiner for his contributions. The research leading to these results has received funding from the European Research Council under European Community's Seventh Framework Programme (FP7/2007-2013 Grant Agreement no.341197).


\begin{thebibliography}{99}

\bibitem{Anderson}
P.W. Anderson, Phys. Rev.{\bf 109}, 1492 (1958).

\bibitem{Abrahams}
E. Abrahams (Ed.), \textit{50 years of Anderson Localization}, World Scientific (vol. 26, 2010).

\bibitem{Fleishman}
L. Fleishman and P.W. Anderson, Phys. Rev. B {\bf 21}, 2366 (1980).

\bibitem{Altshuler}
B.L. Altshuler, Yu. Gefen, A. Kamenev, and L.S. Levitov, Phys. Rev. Lett. {\bf 78}, 2803 (1997).

\bibitem{Fisher} M.P.A. Fisher, P.B. Weichman, G. Grinstein, and D.S. Fisher, Phys. Rev. B {\bf 40}, 546 (1989).

\bibitem{Basko}
D.M. Basko, I.L. Aleiner, and B.L. Altshuler, Annals of Physics {\bf 321}, 1126 (2006).

\bibitem{Billy}
J. Billy, et al., Nature {\bf 453}, 891, (2008).

\bibitem{Roati}
G. Roati, et al., Nature {\bf 453}, 895, (2008).

\bibitem{Sanchez-Palencia}
L. Sanchez-Palencia and M. Lewenstein, Nature Physics {\bf 6}, 87 (2010). 

\bibitem{Tanzi}
L. Tanzi, et al., Phys. Rev. Lett. {\bf 111}, 115301 (2013).

\bibitem{Schreiber}
M. Schreiber, et al., Science {\bf 349}, 842 (2015); P. Bordia, et al., Phys. Rev. Lett. {\bf 116}, 140401 (2016).

\bibitem{Choi}
J. Choi, et al., Science {\bf 352}, 1547 (2016).

\bibitem{Ber1971} V. Berezinskii, Sov. Phys. JETP {\bf 32}, 493 (1971).

\bibitem{Kosterlitz}
J.M. Kosterlitz and D.J. Thouless, Journ. of Phys. C {\bf 6}, 1181 (1973).

\bibitem{Falco}
G.M. Falco, T. Nattermann and V.L. Pokrovsky, Phys. Rev. B {\bf 80}, 104515 (2009).

\bibitem{Carleo}
G. Carleo, G. Boeris, M. Holzmann, and L. Sanchez-Palencia, Phys. Rev. Lett. {\bf 111}, 050406 (2013).

\bibitem{Nandkishore}
R. Nandkishore, Phys. Rev. B {\bf 90}, 184204 (2014).

\bibitem{note1} A preliminary discussion of this phase diagram and the proof of equation (1) are contained in: I.L. Aleiner, B.L. Altshuler, and G.V. Shlyapnikov, arXiv:0910.4534.

\bibitem{Lifshitz}
I.M. Lifshitz, Sov. Phys. Usp. {\bf 7}, 549 (1965).

\bibitem{Zittartz}
J. Zittartz and J.S. Langer, Phys. Rev. {\bf 148}, 741(1966)

\bibitem{Lee}
P.A. Lee and T.V. Ramakrishnan, Rev. Mod. Phys. {\bf 57}, 287 (1985).

\bibitem{Manai}
I. Manai, et al.,  Phys. Rev. Lett. {\bf 115}, 240603 (2015) .

\bibitem{Aleiner}
I.L. Aleiner, B.L. Altshuler and G.V. Shlyapnikov, Nat. Phys. {\bf 6}, 900 (2010).

\bibitem{Michal}
V.P. Michal, B.L. Altshuler and G.V. Shlyapnikov, Phys. Rev. Lett. {\bf 113}, 045304 (2014).

\bibitem{walraven} O.J. Luiten, M.W. Reynolds, and J.T.M. Walraven, Phys. Rev. A {\bf 53}, 381 (1996).

\bibitem{ketterle} W. Ketterle and N.J. Van Druten, Adv. At. Mol. Opt. Phys. {\bf 37}, 181 (1996). 

\bibitem{chang2016} R. Chang, Q. Bouton, H. Cayla, C. Qu, A. Aspect, C.I. Westbrook, and D. Clement, Phys. Rev. Lett. {\bf 117}, 235303 (2016).

\bibitem{Gornyi}
I.V. Gornyi, A.D. Mirlin, M. M\"uller and D.G. Polyakov, Annalen der Physik 2017, 529, 1600365. 

\bibitem{clement} D. Clement, N. Fabbri, L. Fallani, C. Fort, and M. Inguscio, Phys. Rev. Lett. {\bf 102}, 155301 (2009).

\bibitem{ernst} P.T. Ernst, S. Gotze, J.S. Krauser, K. Pyka, D-S. Luhmann, D. Pfannkuche, and K. Sengstock, Nature Physics {\bf 6}, 56 (2010).

\bibitem{Huang}
K. Huang and H.-F. Meng, Phys. Rev. Lett. {\bf 69} 644 (1992).

\bibitem{Meng}
H.-F. Meng, Phys. Rev. B {\bf 49} 1205 (1994).

\bibitem{Nelson}
D.R. Nelson and J.M. Kosterlitz, Phys. Rev. Lett. {\bf 39}, 1201 (1977).

\bibitem{Prokof'ev}
N. Prokof'ev, O. Ruebenacker and B. Svistunov, Phys. Rev. Lett. {\bf 87}, 270402 (2001).

\bibitem{Yukalov}
V. I. Yukalov and R. Graham, Phys. Rev. A {\bf 75}, 023619 (2007).

\bibitem{bktMeasurements}
Z. Hadzibabic et al., Nature {\bf441}, 1118-1121 (2006); P. Kr\"uger, et al., Phys. Rev. Lett. {\bf99}, 040402 (2007); P. Clad\'e, et al., Phys. Rev. Lett. {\bf102}, 170401 (2009).

\bibitem{disorderMeasurements}
M. C. Beeler, M. E. W. Reed, T. Hong, and S. L. Rolston, New J. Phys. {\bf14}, 073024 (2012); B. Allard, et al., Phys. Rev. A {\bf85}, 033602 (2012).

\bibitem{Krinner}
S. Krinner, D. Stadler, J. Meineke, J.-P. Brantut, and T. Esslinger, Phys. Rev. Lett. {\bf110}, 100601 (2013).

\bibitem{boxMeasurements}
A. L. Gaunt, et al., Phys. Rev. Lett. {\bf110}, 200406 (2013); L. Corman, et al, Phys. Rev. Lett. {\bf113}, 135302 (2014); L. Chomaz, et al., Nat. Commun. {\bf6}, 6162 (2015).

\bibitem{Ville}
J. L. Ville, et al., Phys. Rev. A {\bf95} 013632 (2017).

\bibitem{Mathey}
L. Mathey, K.J. G\"unter, J. Dalibard, and A. Polkovnikov, Phys. Rev. A 95 053630 (2017).

\bibitem{BK2014} A. De Luca, B.L. Altshuler, V.E. Kravtsov, and A. Scardicchio, Phys. Rev. Lett. {\bf 113}, 046806 (2014).


\end{thebibliography}
\end{document}


\subsection*{\center Supplemental Material: Finite temperature disordered bosons in two dimensions}

\subsubsection*{Single particle localization in 2D.} Let us consider a weakly bound state with localization length $\zeta$ in a short-range Gaussian random potential $U(x)$, with correlation length $\sigma$ and amplitude $U_0$. Under the condition $\zeta \gg\sigma$, the kinetic energy is given by:
\begin{equation}
K\simeq \frac{\hbar^2}{2m\zeta^2}.
\end{equation}
The potential energy contribution is computed noting that the contribution of a well in 2D is $\sim U_0\sigma^2/\zeta^2$ and multiplying it by $\zeta/\sigma$, which is the square root of the number of wells on the length scale $\zeta$.\\
We now minimize the total energy with respect to $\zeta$:
\begin{equation}
\frac{\partial E} {\partial \zeta}= -\frac{\hbar^2}{m\zeta^3}+U_0\frac{\sigma}{\zeta^2}=0, 
\end{equation}
and obtain a characteristic length $\zeta_\ast \sim \hbar^2/mU_0\sigma$. This gives a characteristic energy $\epsilon_\ast\sim mU_0^2\sigma^2/\hbar^2\ll U_0$.
It is convenient to define the characteristic energy $\epsilon_\ast$ and length $\zeta_\ast$ as:
\begin{equation}
\epsilon_\ast=\frac{ mU_0^2\sigma^2}{\pi\hbar^2}; \qquad \zeta_\ast = \sqrt{\frac{2e^2}{\pi}}\frac{\hbar^2}{mU_0\sigma}.
\end{equation}
Then, the single-particle localization length in two dimensions at $\epsilon>\epsilon_\ast$ \cite{Lee} can be written in the form:
\begin{equation}
\zeta(\epsilon)=\frac{\zeta_\ast}{e} \sqrt{\frac{\epsilon}{\epsilon_\ast}}\; \exp{\frac{\epsilon}{ \epsilon_\ast}}
\end{equation}
so that $\zeta(\epsilon_\ast)=\zeta_\ast$.

\subsubsection*{Temperature dependence of the MBLDT.}
\begin{figure}
    \centering
        \includegraphics[width=0.49\textwidth]{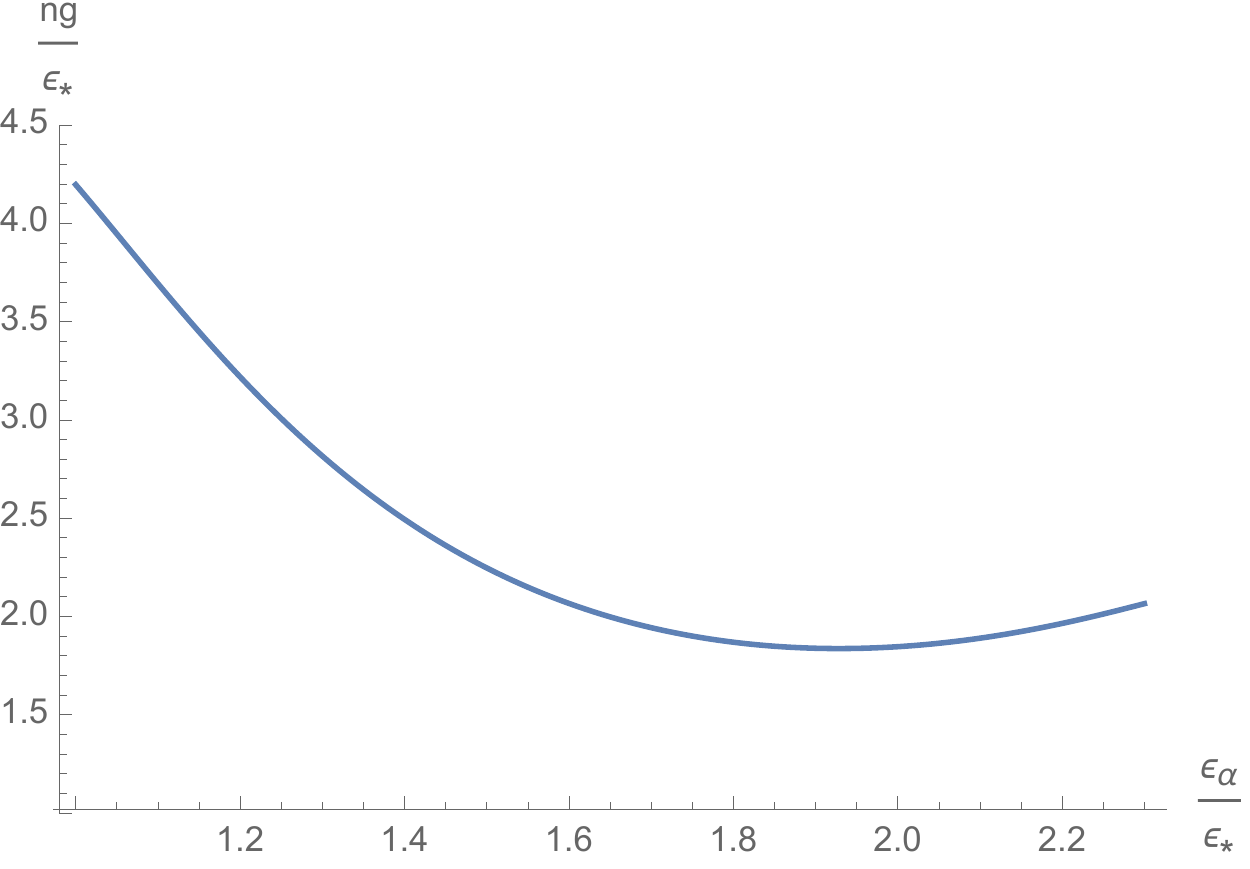}
        \caption{The dependence of the critical coupling $ng$ on $\epsilon_\alpha$ at zero temperature for $C=1$. The minimum is at $\epsilon_\alpha=1.93\epsilon_\ast$.}
        \label{fig:T0}
\end{figure}

As is already said in the main text, at zero temperature on the insulator side the occupation number is given by
\begin{equation}
N_\epsilon = \frac{\zeta^2(\epsilon) \left(\mu-\epsilon\right)}{g}\Theta(\mu-\epsilon),
\end{equation}
where $\Theta$ is the theta-function. Then Eq. (11) of the main text becomes:
\begin{equation}
n= \int_{-\epsilon_\ast}^{\epsilon_\ast} \rho_0 \frac{\zeta_\ast^2 \left(\mu-\epsilon\right)}{g} d\epsilon +\int_{\epsilon_\ast}^\mu \rho_0 \frac{\zeta^2(\epsilon) \left(\mu-\epsilon\right)}{g} d\epsilon,
\end{equation}
and it yields
\begin{equation}\label{eq:normT0}
\frac{ng}{\epsilon_\ast}=f(\mu) \equiv \frac{\exp{(2\mu/\epsilon_\ast)}}{4\pi^3}(\mu/\epsilon_\ast-1)+\frac{e^2}{4\pi^3}(7\mu/\epsilon_\ast+1).
\end{equation}
At the same time, Eq. (9) of the main text gives:
\begin{equation}\label{eq:MBLT0}
g\rho_0^2 \frac{1}{\zeta^2(\epsilon_\alpha)} \int_{-|\epsilon_\ast|}^{\epsilon_\ast} \frac{\zeta_\ast^2 \left(\mu-\epsilon\right)}{g} \zeta_\ast^4 d\epsilon + g\rho_0^2 \frac{1}{\zeta^2(\epsilon_\alpha)} \int_{\epsilon_\ast}^{\epsilon_\alpha} \frac{\zeta^2(\epsilon) \left(\mu-\epsilon\right)}{g} \zeta^4(\epsilon) d\epsilon+ g\rho_0^2 \zeta^2(\epsilon_\alpha) \int_{\epsilon_\alpha}^\mu \frac{\zeta^2(\epsilon) \left(\mu-\epsilon\right)}{g}d\epsilon=C
\end{equation}
where $C$ is a coefficient of order 1. The resulting relation between $\mu$ and $\epsilon_\alpha$ is:
\begin{align*}
F(\epsilon_\alpha,\mu)&\equiv\frac{e^4}{\pi^6}\Bigg( \frac{\zeta(\epsilon_\alpha)^4}{\zeta_*^4} \bigg( \frac{\epsilon_\alpha}{\epsilon_*}\left( \frac{1}{3}-\frac{7\epsilon_*}{18\epsilon_\alpha} +\frac{7\epsilon_*^2}{36\epsilon_\alpha^2} +\frac{\epsilon_*^3}{54\epsilon_\alpha^3} -\frac{\epsilon_*^4}{324\epsilon_\alpha^4} \right) + \frac{\mu}{\epsilon_*}\left( -\frac{1}{3}+\frac{\epsilon_*}{6\epsilon_\alpha}+ \frac{\epsilon_*^2}{36\epsilon_\alpha^2}- \frac{\epsilon_*^3}{216\epsilon_\alpha^3} \right)  \bigg) \Bigg.\\[1em]
&\qquad\qquad\qquad\qquad+\Bigg. \frac{\zeta(\epsilon_\alpha)^2\zeta(\mu)^2}{4\; \zeta_*^4}  \left( 1- \frac{\epsilon_*}{\mu} \right) +\frac{\zeta_*^2}{\zeta(\epsilon_\alpha)^2} \left( \frac{409\mu}{216\epsilon_*} +\frac{31}{324}\right) \Bigg)=C.
\numberthis\label{eq:bigFT0}
\end{align*}
We found numerically from equations \eqref{eq:normT0} and \eqref{eq:bigFT0}, with $C=1$, that the coupling strength is minimal for $\epsilon_{\alpha0}=1.93\epsilon_\ast$. It is equal to $ng_{c0}=1.84\epsilon_\ast$, and the related chemical potential is $\mu_0=2.23\epsilon_\ast$. This is seen from the obtained dependence of $ng_{c0}$ on $\epsilon_\alpha$ (Fig. \ref{fig:T0}).

\begin{figure}[h!]
    \centering
        \includegraphics[width=0.49\textwidth]{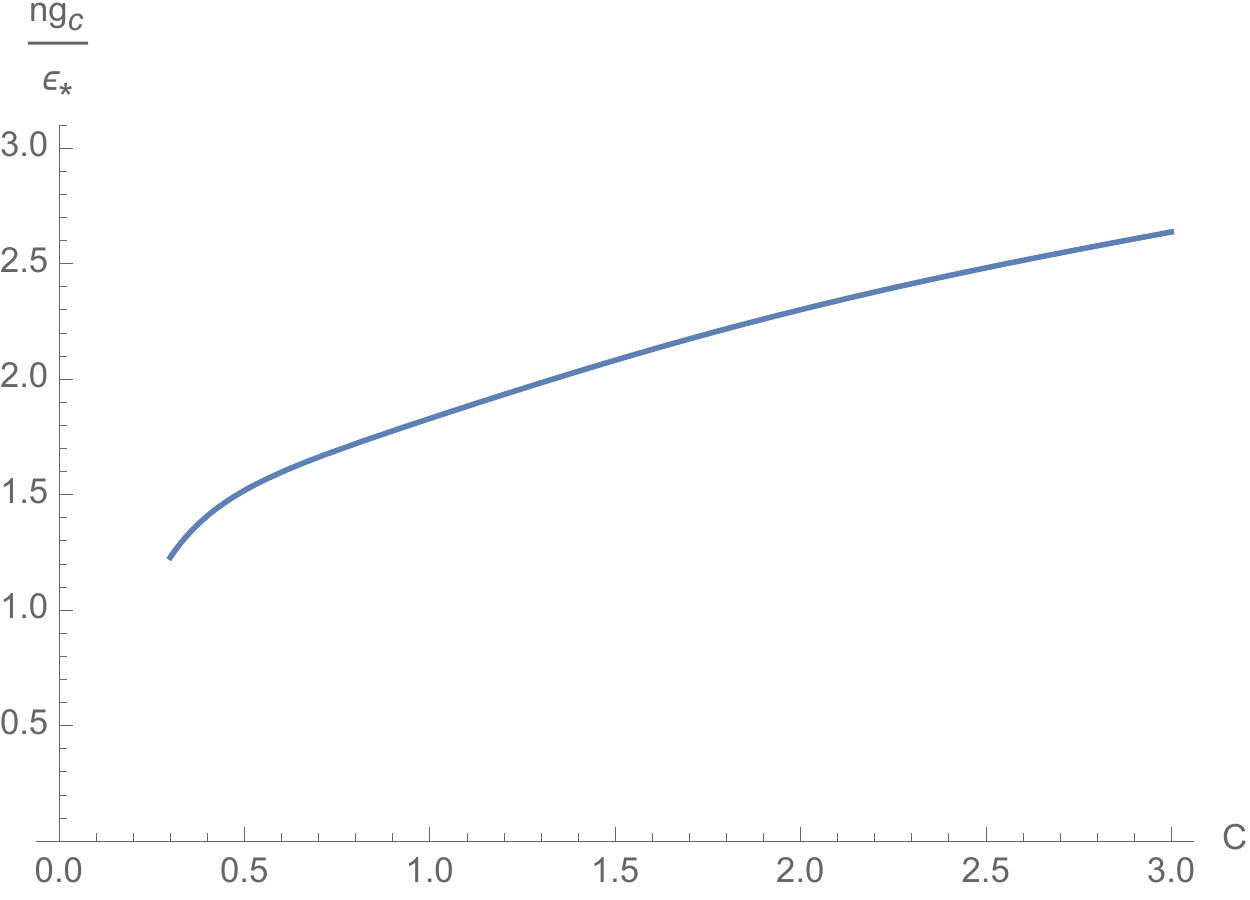}
        \caption{The critical coupling $ng_c0/\epsilon_*$  as a function of the constant $C$.}
        \label{fig:diffC}
\end{figure}

It is instructive to look what happens for different values of the constant $C$ in equation \eqref{eq:bigFT0}. Figure \ref{fig:diffC} shows the zero-temperature critical coupling $ng_{c0}$ obtained for values of $C$ between 0.3 and 3.

At temperatures $T\ll\epsilon_\ast^2/T_d$, the corrections to the zero temperature result are negligible. For temperatures $\epsilon_\ast^2/T_d\ll T\ll \epsilon_\ast$, using equations (11) and (12) of the main text we have the following relations for the occupation numbers:
\begin{equation}\label{eq:occNumber}
N_\epsilon = \begin{cases}
\dfrac{\zeta^2(\epsilon) \left(\mu-\epsilon\right)}{g}+\dfrac{T}{\mu-\epsilon}; & \qquad -\epsilon_*<\epsilon<\mu-\delta\\[1em]
\sim \dfrac{\zeta^2(\mu)}{2g} \left(\mu-\epsilon+\sqrt{\strut(\epsilon-\mu)^2+4Tg/\zeta^2(\mu)}\;\right);& \qquad\mu-\delta<\epsilon<\mu+\delta\\[1em]
(e^{(\epsilon-\mu)/T}-1)^{-1}, & \qquad\mu+\delta<\epsilon
\end{cases}
\end{equation}
where $\delta$ is a small quantity such that $\sqrt{\mathstrut Tg/\zeta^2(\mu)}\ll\delta\ll T$. \\
Eq. (10) of the main text gives
\begin{equation}
\frac{ng_c}{\epsilon_\ast}=f(\mu)+\frac{T}{T_d}\frac{ng_{c0}}{\epsilon_\ast}\ln\left(\frac{(\mu_0+\epsilon_\ast)\zeta^2(\mu_0)}{g_{c0}}\right).
\end{equation}
As the correction to the chemical potential should be small, we can expand $f(\mu)$ near $\mu_0$ and obtain:
\begin{equation}\label{eq:normFiniteT}
\frac{ng_c}{\epsilon_\ast}=f(\mu_0)+\frac{(\mu-\mu_0)}{\epsilon_\ast} f'_\mu(\mu_0)+\frac{T}{T_d}\frac{ng_{c0}}{\epsilon_\ast}\ln\left(\frac{(\mu_0+\epsilon_\ast)\zeta^2(\mu_0)}{g_{c0}}\right).
\end{equation}
Similarly, in the MBLDT criterion at finite temperatures we expand the function $F(\epsilon_\alpha,\mu)$ near $\mu_0$ and $\epsilon_{\alpha0}$, which gives: 
\begin{equation}
F(\epsilon_{\alpha0},\mu_0)+\frac{(\mu-\mu_0)}{\epsilon_\ast}F'_\mu(\epsilon_{\alpha0},\mu_0)+\frac{(\epsilon_\alpha-\epsilon_{\alpha0})}{\epsilon_\ast} F'_{\epsilon_\alpha}(\epsilon_{\alpha0},\mu_0)+\frac{T}{T_d} \frac{ng_{c0}}{\epsilon_\ast}G(\epsilon_{\alpha0},\mu_0,g_{c0})=C
\end{equation}
with  
\begin{align*}
G(\epsilon_{\alpha0},\mu_0,g_{c0})\equiv&\frac{e^2}{\pi^3}\Bigg( \frac{\zeta(\alpha_0)^2}{\zeta_*^2}\left(\frac{\epsilon_*^2}{16 \epsilon_{\alpha0}^2} \left(1-\frac{4\epsilon_{\alpha0}}{\epsilon_*} -4\frac{\mu_0}{\epsilon_\ast}\right) +\ln \left(\frac{(\mu_0-\epsilon_{\alpha0})\;\zeta^2(\mu_0)}{g_{c0}} \right)\right)\Bigg.\\
 &+\frac{\zeta(\mu_0)^4}{\zeta(\epsilon_{\alpha_0})^2\zeta_*^2} \left(\text{Ei}\left(4-4\frac{\mu_0}{\epsilon_\ast}\right)-\text{Ei}\left(4\frac{\epsilon_{\alpha0}}{\epsilon_\ast}-4\frac{\mu_0}{\epsilon_\ast}\right)\right)+\frac{\zeta_*^2}{\zeta(\epsilon_{\alpha_0})^2}\left(\frac{3}{16}+\frac{\mu_0}{4\epsilon_\ast} +\ln\left(\frac{\mu+\epsilon_*}{\mu-\epsilon_*}\right) \right) \Bigg), \numberthis
\end{align*}
where $\text{Ei}(x)$ is the exponential integral function. We have inserted the zero-temperature values of $\mu_0$ and $g_{c0}$ in the temperature correction, as these corrections are of order $T/T_d$. By construction we have $F'_{\epsilon_\alpha}(\epsilon_{\alpha0},\mu_0) =0$ and $F(\epsilon_{\alpha0},\mu_0)=C$. This results in an expression for $\mu-\mu_0$ which we substitute into Eq. \eqref{eq:normFiniteT}. We then obtain:
\begin{equation}
\frac{ng_c}{\epsilon_\ast}=\frac{ng_{c0}}{\epsilon_\ast}\Bigg(1-\frac{T}{T_d}\Bigg(\frac{f'_\mu(\mu_0)}{F'_\mu(\epsilon_{\alpha0},\mu_0)}G(\epsilon_{\alpha0},\mu_0,g_{c0})-\ln\left(\frac{(\mu_0+\epsilon_\ast)\zeta^2(\mu_0)}{g_{c0}}\right) \Bigg).
\end{equation}
\begin{figure}[h!]
    \centering
        \includegraphics[width=0.48\textwidth]{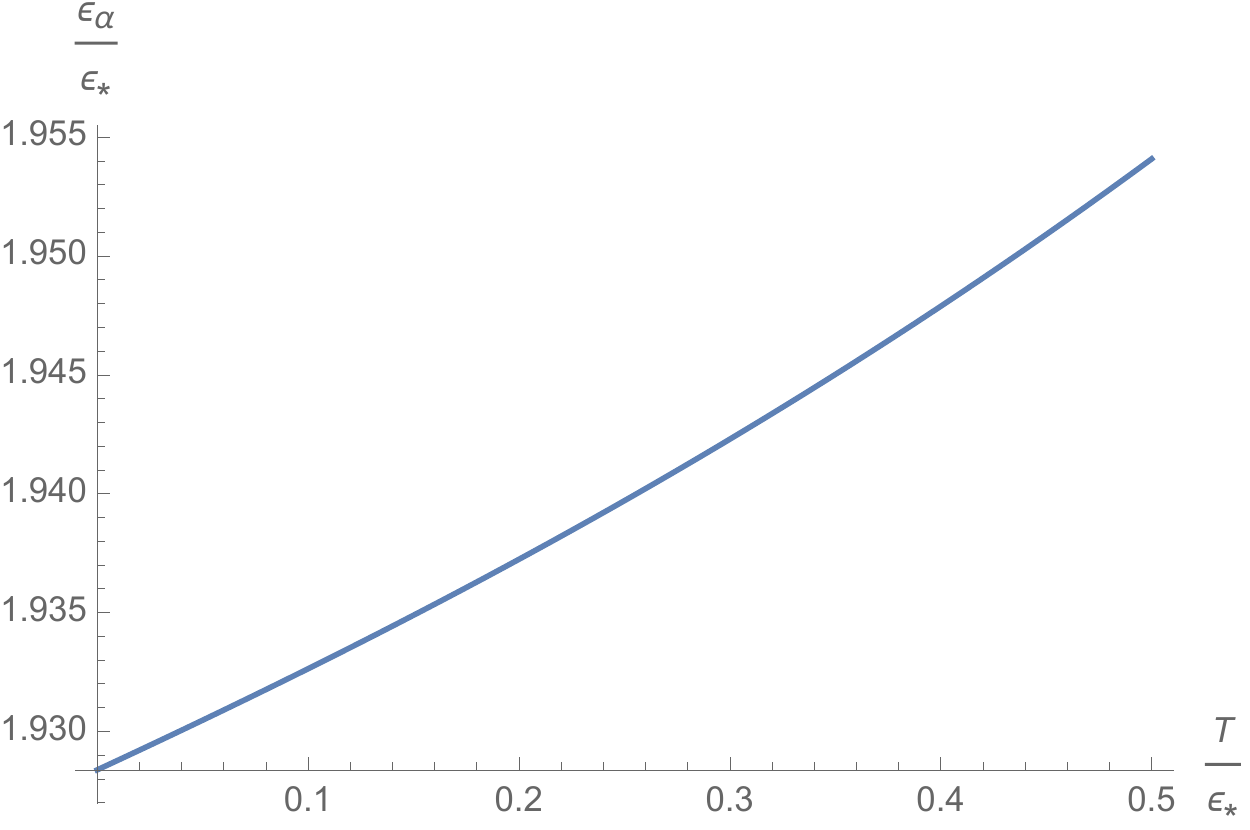}
        \caption{The temperature dependence of the value of $\epsilon_\alpha$ that minimizes $g_c$ at $C=1$. We used here $T_d/\epsilon_*=50$.}
        \label{fig:smallT}
\end{figure}

By setting $C=1$ and inserting the corresponding values of  $ng_{c0}$, $\mu_{0}$, $\epsilon_{\alpha0}$ in the temperature corrections, we obtain Eq. (15) of the main text. The numerically obtained temperature dependence of $\epsilon_\alpha$ that minimizes $g_c$ is given in Fig. \ref{fig:smallT} for $C=1$, and one sees that the optimal $\epsilon_\alpha$ is very close to $\epsilon_{\alpha0}$. For different values of $C$ the numerical coefficients in Eq. (15) of the main text vary only slightly. 

In the thermodynamic limit, the critical coupling tends to zero when $T\to\epsilon_\ast/2$ from below. We may assume that the chemical potential decreases below $\epsilon_\ast$, so that $\mu\to-|\epsilon_*|$ when $g_c \to 0$, and we expect $\epsilon_\alpha$ to increase. Equation (9) of the main text takes the form:
\begin{equation}
g\rho_0^2 \Bigg(\frac{1}{\zeta^2(\epsilon_\alpha)} \int_{-\epsilon_\ast}^{\epsilon_\ast} N_\epsilon \zeta_\ast^4 d\epsilon + \frac{1}{\zeta^2(\epsilon_\alpha)} \int_{\epsilon_\ast}^{\epsilon_\alpha} \frac{\zeta^4(\epsilon)}{e^{(\epsilon-\mu)/T}-1} d\epsilon+ \zeta^2(\epsilon_\alpha)\int_{\epsilon_\alpha}^\infty \frac{1}{e^{(\epsilon-\mu)/T}-1} d\epsilon\Bigg)\approx C.
\end{equation}
As $\epsilon_\alpha$ is large, we neglect the first integral and calculating the other integrals we keep only the highest power in $\epsilon_\alpha$. The coupling strength $g$ is then minimized at $\epsilon_\alpha=T\epsilon_\ast/(\epsilon_\ast-2T)$. Using $\mu\simeq-|\epsilon_\ast|$ leads to the equation:
\begin{equation}     \label{gclimit}
ng_c\simeq 4(\pi e)^3\frac{T_d}{\epsilon_{\ast}}\left(\frac{\epsilon_{\ast}}{2}-T\right);\,\,\,\,\,T\rightarrow \frac{\epsilon_{\ast}}{2}.
\end{equation}

\begin{figure}[h!]
    \begin{subfigure}[b]{0.45\textwidth}
        \includegraphics[width=\textwidth]{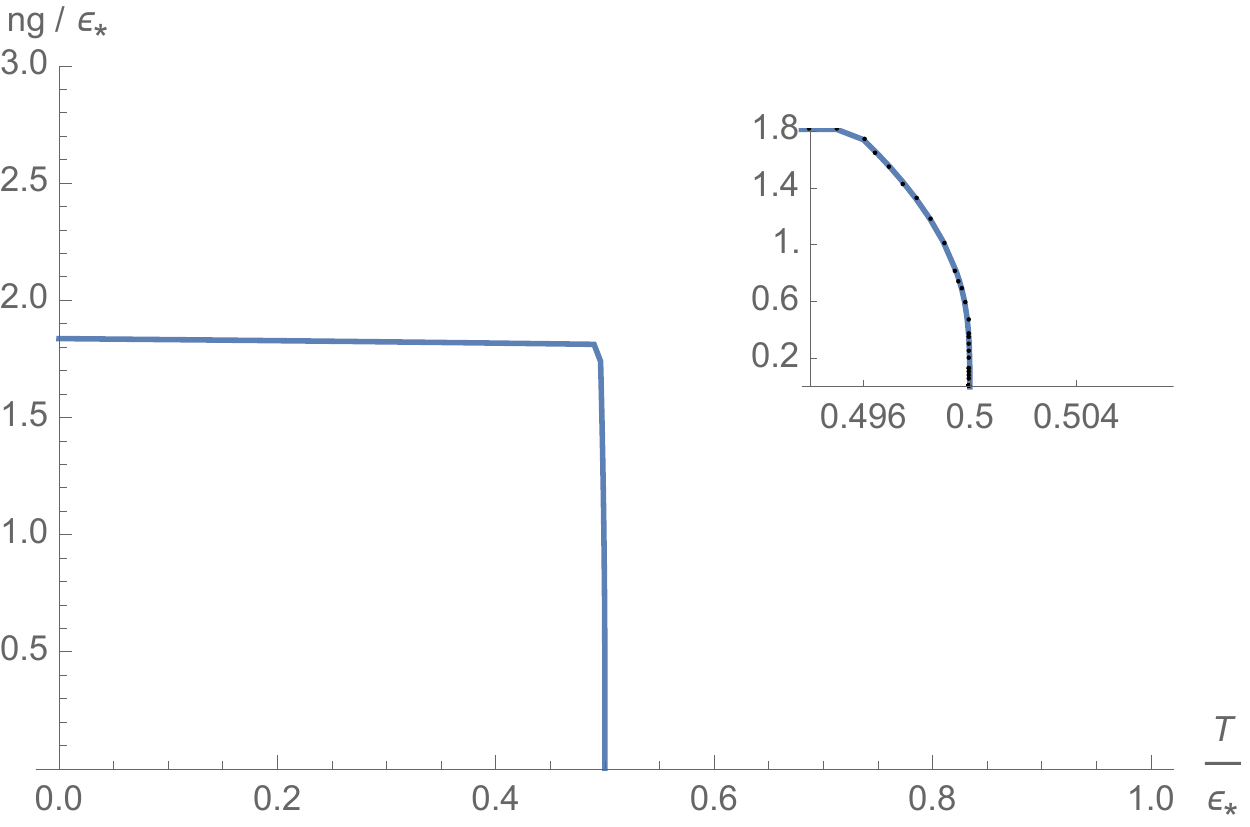}
        \label{fig:ng}
         \caption{\small}
    \end{subfigure}
    \qquad
        \begin{subfigure}[b]{0.45\textwidth}
        \includegraphics[width=\textwidth]{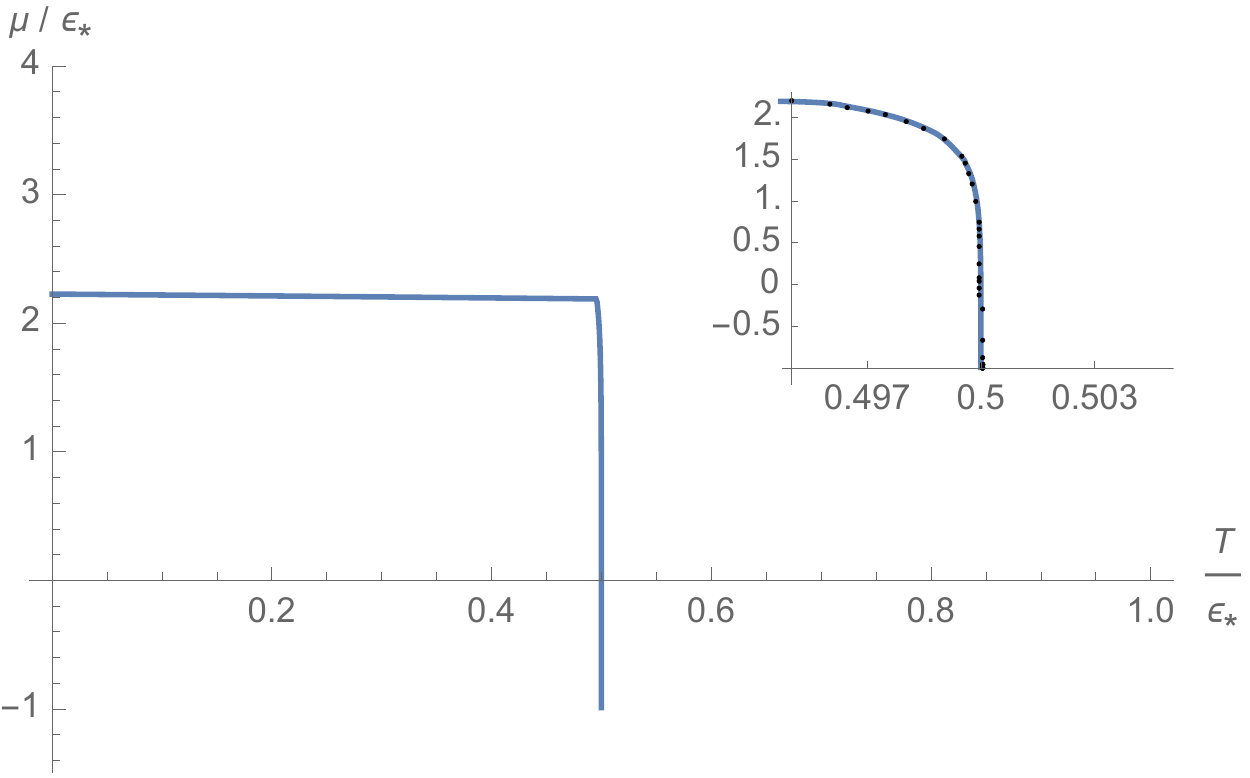}
        \caption{\small}
        \label{fig:mu}
    \end{subfigure}
    \caption{The critical coupling $ng_{c0}$ in (a) and chemical potential $\mu$ in (b) versus temperature for $C=1$. In both insets we see that the derivative is always finite and becomes very large when we approach $T=\epsilon_\ast/2$.}    
    \label{fig:TdepDiffC}
\end{figure}

The results of exact numerics for $g_c(T)$ and $\mu(T)$ in the thermodynamic limit are shown in Figure \ref{fig:TdepDiffC}  $C=1$. We can see that the values of both $g_c(T)$ and $\mu(T)$ are almost constant until we get to the vicinity of $T=\epsilon_\ast/2$. Here they both sharply drop with a finite but large derivative, as shown in the insets. This happens irrespective of the value of $C$. In the small intermediate region of $T$ given by the insets, where the analytical approach is not possible, we numerically solved equations (9) and (10) of the main text, with the occupation numbers given by equation \eqref{eq:occNumber} in the Supplemental Material.  

\subsubsection*{MBLDT for the truncated distribution function}

In the main text we argued that in realistic systems the distribution function $N_\epsilon$ is truncated at an energy $\epsilon_b$. We take $\epsilon_b=1.21ng+\eta T$, where $\eta$ ranges from 5 to 8. This means that the MBLDT criterion reads
\begin{equation}
C=g\rho_0^2\left( \frac{1}{\zeta^2(\epsilon_\alpha)}\int_{-\epsilon_*}^{\epsilon_\alpha} N_\epsilon \zeta^4(\epsilon) d\epsilon + \zeta^2(\epsilon_\alpha)\int_{\epsilon_\alpha}^{\epsilon_b} N_\epsilon \right).
\end{equation}
At $T\ll\epsilon_\ast$ the truncation practically does not influence the results. However, at higher temperatures the influence is crucial. Considering $T>\epsilon_\ast/2$ and setting $\epsilon_\alpha \to \epsilon_b$, with $\mu+\delta<\epsilon_*$, one gets
\begin{align}
&C=g\rho_0^2  \frac{1}{\zeta^2(\epsilon_b)} \left(\int_{-\epsilon_*}^{\mu+\delta} \frac{\zeta_*^2}{2g}\left(\mu-\epsilon+\sqrt{(\mu-\epsilon)^2+4Tg/\zeta^2_*}\right) \zeta^4_* d\epsilon \right. \nonumber
\\&\qquad\qquad\qquad\left.+\int^{\epsilon_*}_{\mu+\delta} \frac{1}{e^{(\epsilon-\mu)/T}-1} \zeta^4_* d\epsilon + \int_{\epsilon_*}^{\epsilon_b} \frac{1}{e^{(\epsilon-\mu)/T}-1} \zeta^4 (\epsilon) d\epsilon   \right).\label{MBLDThighT}
\end{align}
The last integral dominates and gives
\begin{equation}\label{MBLDTtrunc}
C=\pi^{-3}\frac{ng}{\epsilon_*}\frac{\epsilon_b}{\epsilon_*}\frac{T}{T_d}\frac{1}{4T/\epsilon_*-1}\exp\left\{\epsilon_b\left(\frac{2}{\epsilon_*}-\frac{1}{T}\right)+\frac{\mu}{T} \right\}.
\end{equation}
From Eq. (11) of the main text one has:
\begin{equation}
\mu\simeq \frac{\pi^{3/2}}{\sqrt{2e^2}} ng - \left(1-\frac{\pi^{3/2}}{\sqrt{2e^2}}\right) \epsilon_*.
\end{equation}
For the critical coupling Eq. \eqref{MBLDTtrunc} then yields:
\begin{equation}
ng_c\simeq C \sqrt{\frac{2e^2}{\pi^3}} \; T \; W \left(\sqrt{\frac{\pi ^{9}}{2e^2}} \frac{\epsilon _*^2}{T^2}\frac{T_d}{\epsilon_b} \left(\frac{4 T}{\epsilon _*}-1\right) e^{\epsilon _b \left(\frac{1}{T}-\frac{2}{\epsilon _*}\right)}\right)\label{ngbarrier},
  \end{equation}
where $W(x)$ is the (main branch) Lambert $W$-function defined as $x=W(xe^x)$. For $T>\epsilon_*/2$, the argument of $W(x)$ is small. We then use the approximation $W(x)\approx x $ for small $x$. This gives:
\begin{equation}
ng_c\simeq C \; \pi^3 \epsilon_* \frac{T_d}{\epsilon_b} \left(4-\frac{\epsilon _*}{T}\right) e^{\epsilon _b \left(\frac{1}{T}-\frac{2}{\epsilon _*}\right)}.
  \end{equation}
The term $ (4 -\epsilon_*/T)$ takes values ranging from 2 to 4 as $T$ is increased above $\epsilon_*/2$. We take $ (4 -\epsilon_*/T)\approx 2$ for simplicity, which yields:
\begin{equation}
\epsilon_*^{MBL} = \frac{2\epsilon_b}{W\left( 4\pi^3\dfrac{T_d e^{\epsilon_b/T}}{ng} \right)}.
\end{equation}
The argument of $W(x)$ is now large and we can use $W(x)\approx (\ln x- \ln\ln x)$ for large $x$ to get Eq. (16) of the main text, using $C=1$. The comparison of this analytical expression with exact numerics is given in Figure 5, where we used $\epsilon_b=1.21ng+5T$. In Fig. 6 we show the same quantities as in Fig. 5, but for the truncation of the energy distribution function at $\epsilon_b=1.21ng+8T$. As one sees, the increase of $\beta$ from 5 to 8 does not significantly change the MBLDT transition line.
\begin{figure}[h!]
    \begin{subfigure}[b]{0.4\textwidth}
        \includegraphics[width=\textwidth]{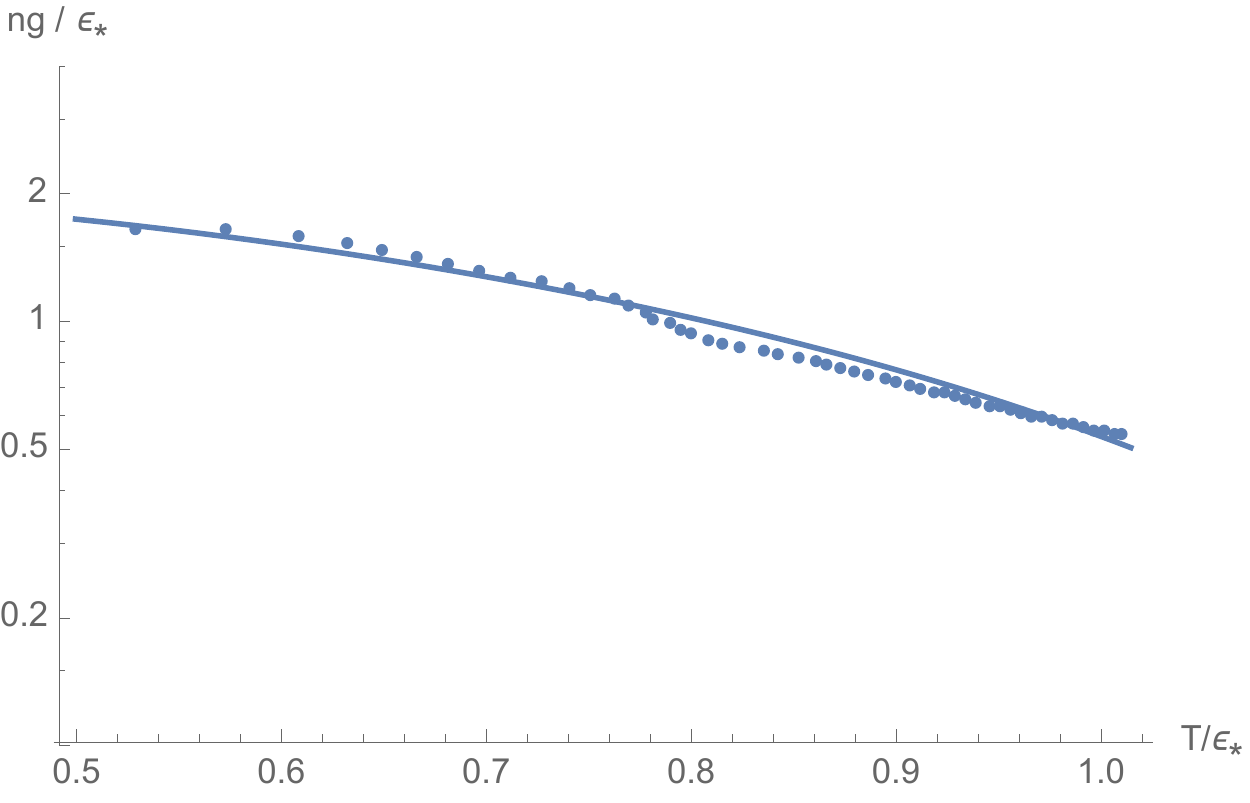}
        \label{fig:ng}
         \caption{\small}
    \end{subfigure}
    \qquad
        \begin{subfigure}[b]{0.4\textwidth}
        \includegraphics[width=\textwidth]{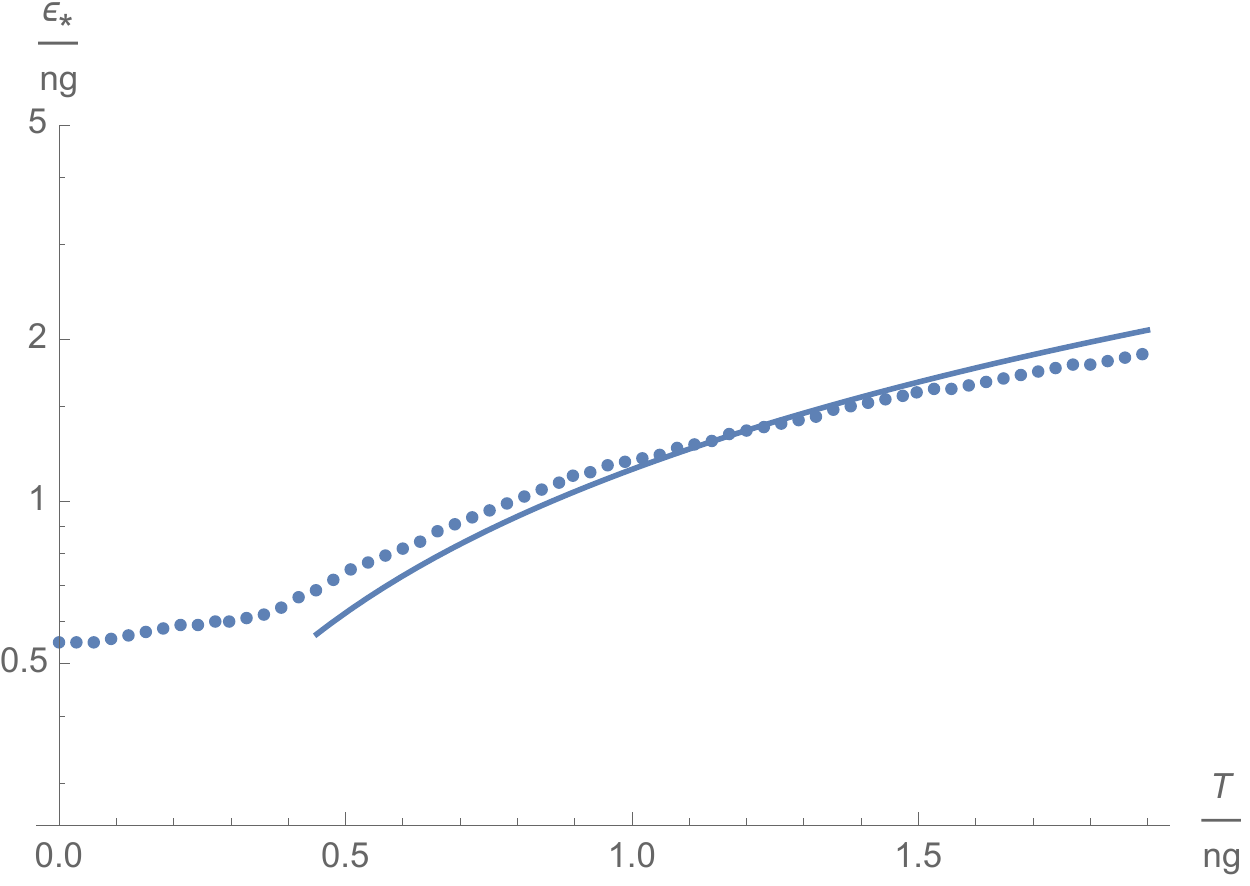}
        \caption{\small}
        \label{fig:mu}
    \end{subfigure}
    \caption{Comparison of numerical results with those from analytical expressions for $\epsilon_b = 1.21ng+5T$. The dots are the results of numerically solving Eq.(19), and the solid curve is given by Eq. \eqref{ngbarrier} in (a) and by Eq. (16) of the main text in (b). We used $T_d/\epsilon_*=20$ in (a) and $T_d/ng=11$ in (b).}    
      \label{fig:TbiggerEps/2}
\end{figure} 

\begin{figure}[h!]
    \begin{subfigure}[b]{0.4\textwidth}
        \includegraphics[width=\textwidth]{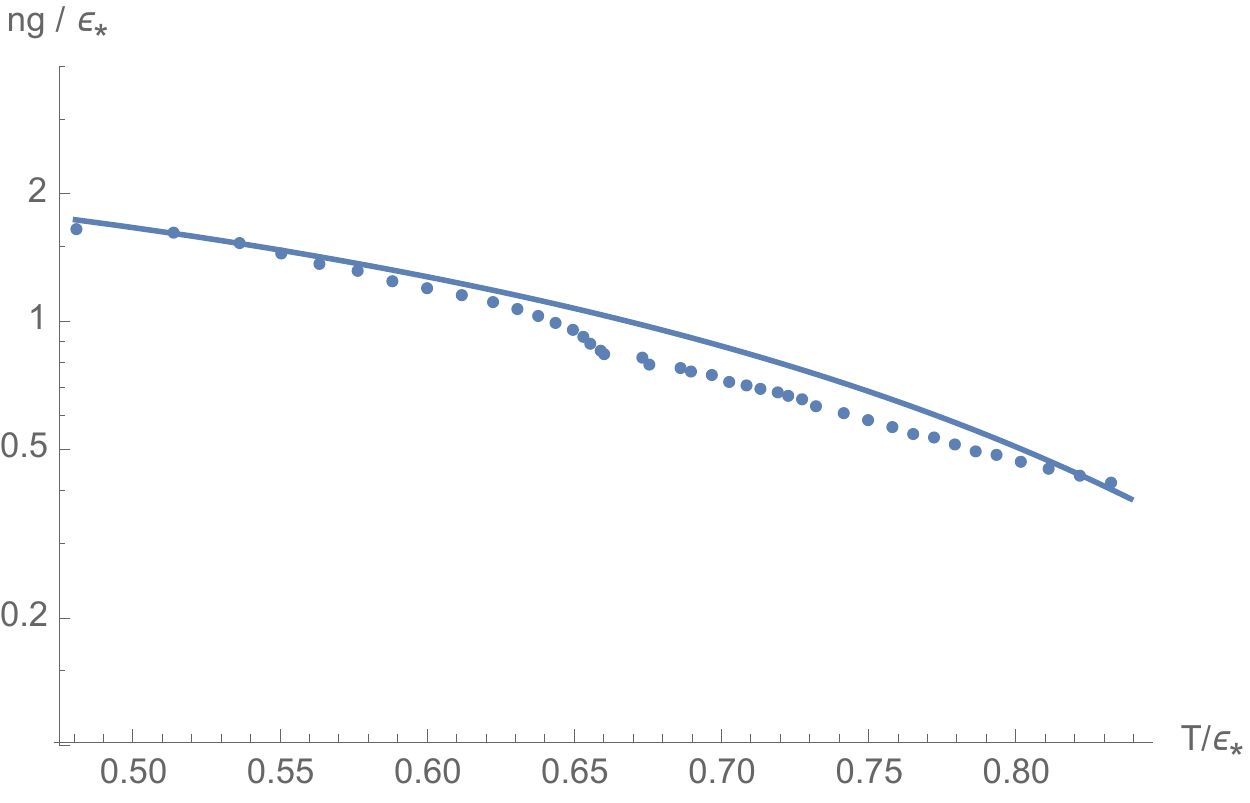}
        \label{fig:ng}
         \caption{\small}
    \end{subfigure}
    \qquad
        \begin{subfigure}[b]{0.4\textwidth}
        \includegraphics[width=\textwidth]{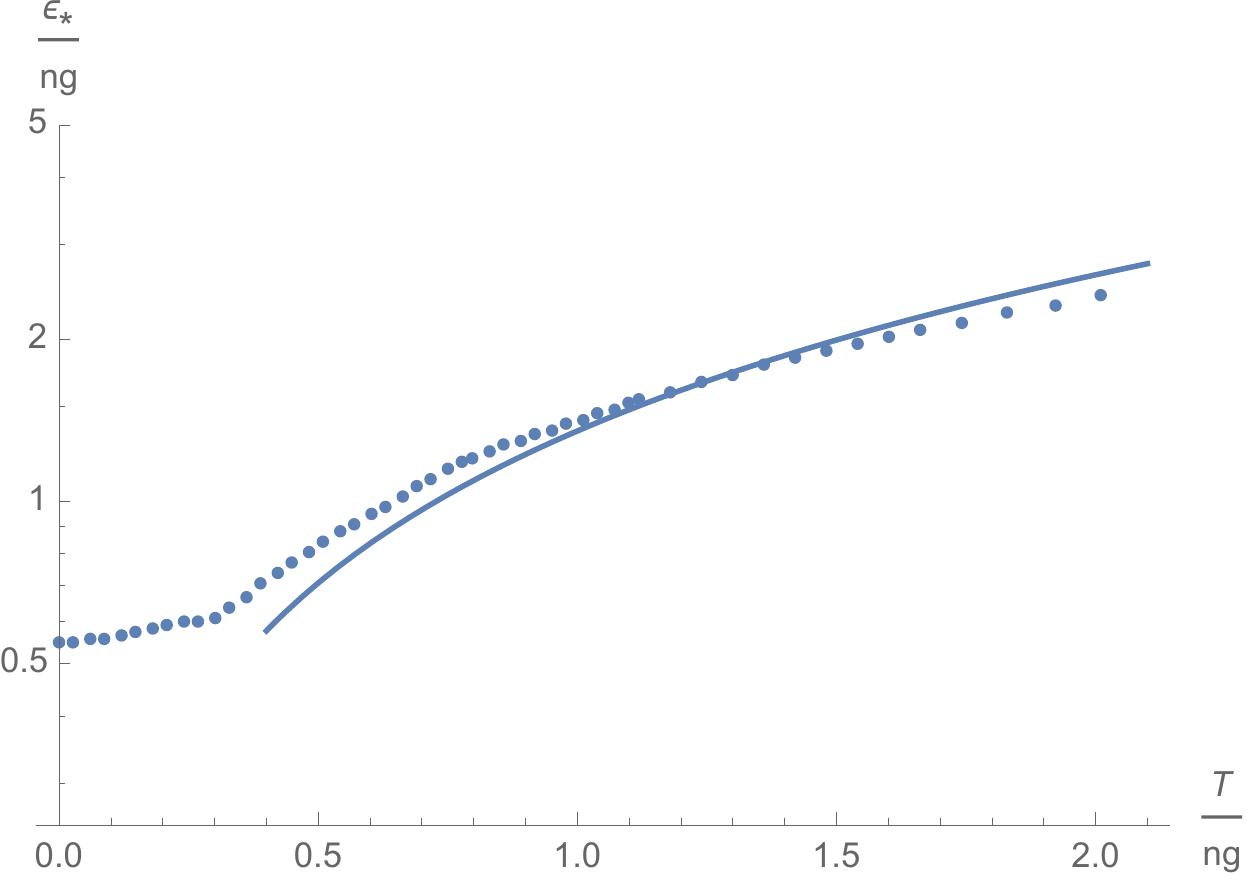}
        \caption{\small}
        \label{fig:mu}
    \end{subfigure}
    \caption{The same as in Fig.5, but for the truncation of the energy distribution function at $\epsilon_b = 1.21ng+8T$.}    
      \label{fig:TbiggerEps/28T}
\end{figure} 

\subsubsection*{Tricritcal point at $T=0$}

This subsection is dedicated to ruling out a direct transition from the insulator to superfluid phase. This transition would mean that one has a phase diagram either like Fig. (a) or Fig. (b).

\begin{figure}[h!]
\center
\begin{subfigure}{0.4\textwidth}
\begin{tikzpicture}
\draw[gray, thick,->] (0,0) -- (0,3) node[above] {$\epsilon_\ast$}; 
\draw[gray, thick,->] (0,0) -- (4.5,0) node[right] {$T$};
\draw [thick,blue] (1.75,1.75) to[out=15,in=-115] (3.8,2.8); 
\draw [thick,black] (0,2) to[out=-5,in=135] (4,0.5); 
\draw (1.5,0.9) node [above,left] {SF};
\draw (3.8,1.7) node [above,left] {NF};
\draw (1.5,2.5) node [above,left] {INS};
\end{tikzpicture}
\label{fig:SI1}
\caption{\small}
\end{subfigure}
\begin{subfigure}{0.4\textwidth}
\begin{tikzpicture}
\draw[gray, thick,->] (0,0) -- (0,3) node[above] {$\epsilon_\ast$}; 
\draw[gray, thick,->] (0,0) -- (4.5,0) node[right] {$T$};
\draw [thick,blue] (0,1.4) to[out=5,in=-125] (3.8,2.8); 
\draw [thick,black] (1.8,1.6) to[out=-5,in=135] (4,0.5); 
\draw (1.5,0.9) node [above,left] {SF};
\draw (3.8,1.7) node [above,left] {NF};
\draw (1.5,2.5) node [above,left] {INS};
\end{tikzpicture}
\label{fig:SI2}
\caption{\small}
\end{subfigure}
\end{figure}

Fig. (a) is ruled out as follows. At the critical zero-temperature disorder and arbitrarily small finite temperatures, the system is unstable with respect to delocalization. This conclusion is supported by the quantitative analysis given in the text, where corrections to the zero-temperature value $\epsilon_*^{MBL}(0)$ were found to be positive. The physical picture is that although excitations in the insulator at $T=0$ are always localized, their localization length at the critical disorder strength $\epsilon_*^{MBL}(0)$ can be arbitrarily large. Therefore, at any fixed disorder $\epsilon_*>\epsilon_*^{MBL}(0)$ the elementary excitations undergo many-body delocalization with increasing temperature. The critical temperature tends to zero as the localization length diverges, i.e. at arbitrary low finite temperatures there will be a range of disorder strengths corresponding to a normal fluid. 
Fig. (b) is ruled out by the following arguments. The zero-temperature insulator can be viewed as a composition of superfluid lakes with uncorrelated phases, which are separated from each other by a certain distance. Tunneling of particles between the lakes increases with decreasing the disorder strength, and at a critical disorder it establishes the phase coherence between the lakes, so that the whole system becomes superfluid. Consider a single lake at the disorder strength $\epsilon_*^{MBL}(0)+0$ and slightly increase the temperature. Then a certain fraction of particles in the lake will become non-superfluid. Assuming slowly varying density fluctuations, such that the Bogoliubov theory works, the non-superfluid fraction within the lake turns out to be 
$$\frac{n'}{n}=3\zeta(3)\frac{T^3}{T_d n^2g^2},$$ 
where $\zeta(3)$ is the Riemann zeta-function, $g$ is the coupling strength of the interparticle interaction, $n$ is the density, and $T_d$ the degeneracy temperature. A decrease of the superfluid density $n$ (which is equivalent to decreasing the coupling strength $g$) reduces the probability of tunneling between neighbouring lakes, which behaves as (see e.g., \cite{Falco})
$$t\sim\exp\left(-\sqrt{\epsilon_*^{MBL}(0)/ng}\right),$$  
and the latter is unable to establish phase coherence between the lakes.